# Generations of polygonal soliton clusters and fundamental solitons by radially-azimuthally phase-modulated necklace-ring beams in dissipative systems

Yingji He,[1]* Dumitru Mihalache,[2,3] Boris A. Malomed,[4] Yunli Qiu,[1] Zhanxu Chen,[1]

and Yifang Li [1]

[1]*School of Electronics and Information, Guangdong Polytechnic Normal University,*

*510665 Guangzhou, China*

[2]*Horia Hulubei National Institute for Physics and Nuclear Engineering, P.O.B. MG-6,*

*077125 Magurele-Bucharest, Romania*

[3]*Academy of Romanian Scientists, 54 Splaiul Independentei, 050094 Bucharest,*

*Romania*

[4]*Department of Physical Electronics, School of Electrical Engineering, Faculty of*

*Engineering, Tel Aviv University Tel Aviv 69978, Israel*

* Corresponding author: heyingji8@126.com

**Abstract:** We demonstrate that, in a two-dimensional dissipative medium described by the cubic-quintic (CQ) complex Ginzburg-Landau (CGL) equation with the viscous (spectral-filtering) term, necklace rings carrying a mixed radial-azimuthal phase modulation can evolve into polygonal or quasi-polygonal stable soliton clusters, and into stable fundamental solitons. The outcome of the evolution is controlled by the depth and azimuthal anharmonicity of the phase-modulation profile, or by the radius and number of "beads" in the initial necklace ring. Threshold characteristics of the evolution of the patterns are identified and explained. Parameter regions for the formation of the stable polygonal and quasi-polygonal soliton clusters, and of stable fundamental solitons, are identified. The model with the CQ terms replaced by the full saturable nonlinearity produces essentially the same set of the basic dynamical scenarios; hence this set is a universal one for the CGL models.

**PACS numbers**: 47.54.-r, 42.65.Tg, 42.65.Sf

**Keywords:** spatial solitons; soliton clusters; phase modulations; Ginzburg-Landau equations; cubic-quintic nonlinearity;

## 1. Introduction

Spatial optical solitons in conservative and dissipative media have drawn a great deal of interest in recent years [1-14], due to their potential for applications to all-optical switching, pattern recognition, and parallel data processing [3]. In this context, the possibility of generation of spatial soliton arrays in laser cavities has been investigated in various settings [7,8,12,15].

Many recent works focused on localized complex patterns in conservative models of optical media, different from the simplest nodeless ground-state modes, such as vortex solitons [16], soliton clusters [17,18], dipole-mode structures and their multipole counterparts [19-22], and necklace-ring solitons [23-25]. Dissipative solitons, including vortical and necklace-shaped patterns [26-35], may be stable in the framework of the cubic-quintic (CQ) complex Ginzburg-Landau (CGL) equations, which account for the presence of the inner saturable gain in the physical medium. It is well known that CGL equations represent a broad class of models with applications to superconductivity, nonlinear optics, plasmas, Bose-Einstein condensates, and quantum field theories [36-47], diverse realizations in terms of laser media being especially important [37-47]. In particular, we have recently demonstrated that arrays of dissipative solitons can be induced by means of spatial phase modulations in the two-dimensional (2D) and three-dimensional (3D) CQ CGL models [48, 49].

Quasi-localized dissipative patterns ("spots"), built on top of a finite-amplitude background, and clusters composed of them, have been studied too, in models of pumped bistable optical cavities based on 1D and 2D Swift-Hohenberg equations [50-52]. In the framework of such systems, the general mechanism of the clusterization [50], effects of the zero mode on interactions between the spots and the stability of clusters and periodic structures [51], and the selection of stable clusters in the 1D and 2D settings by means of the Maxwell's rule [52] have been explored in detail. The difference of the models based on the CGL equations is that they describe physical systems with the inner gain, rather than external pump; accordingly, the localized patterns and clusters are generated by the CGL equations as dissipative solitons, without any background.

In this context, necklace-shaped patterns are universal inputs used for the formation of diverse stable structures, as a result of a sufficiently long evolution. In addition to the initial amplitude profile, the outcome of the evolution may be controlled by phase patterns imprinted onto the initial amplitude distributions. Developing this general approach, in this work we study the evolution of 2D necklace-ring beams (NRBs) governed by the CQ CGL equation, under the action of initial radial-azimuthal phase modulations. Simulations demonstrate that, in this case, NRBs may evolve into stable patterns of three distinct generic types: (a) regular polygonal soliton clusters, (b) quasi-polygonal clusters, and (c) fundamental solitons. The outcome of the evolution is determined by the depth and azimuthal anharmonicity of the phase modulation profile, and by the radius and number of "beads" in the initial necklace beam. If the depth of the initial phase-modulation pattern or the necklace's radius exceed certain minimum values, $p_{min}$ and $R_{min}$, respectively, the NRB evolves into stable polygonal or quasi-polygonal soliton clusters; otherwise (in particular, if the phase-modulation is too shallow), the necklace will fuse into a single stable fundamental soliton. On the other hand, we also find that, if the depth of the phase modulation exceeds a critical maximum value, $p_{max}$, the necklace decays due to the strong loss. Thus, this technique can produce a variety of stable polygonal soliton clusters and fundamental solitons in nonlinear optical media with the inner saturable gain, with the obvious potential for experimental realizations.

The model is introduced in Section 2, and basic results obtained by means of systematic simulations are reported in Section 3.

## 2. The model

We consider the CQ CGL equation of the general form [28-35,53], which is written in terms of the laser-cavity model:

$$iu_z + (1/2)\Delta u + |u|^2 u + \nu |u|^4 u = iR[u] + V(x,y)u, \qquad (1)$$

where $\Delta = \partial^2/\partial x^2 + \partial^2/\partial y^2$ is the transverse diffraction operator, $z$ is the propagation distance, and the coefficient in front of the cubic self-focusing term is scaled to be 1. Further, $\nu$ is the quintic self-defocusing coefficient, and the CQ combination of the loss and gain terms is $R[u] = \delta u + \beta \Delta u + \varepsilon |u|^2 u + \mu |u|^4 u$, with $-\delta$ the linear loss coefficient, $-\mu$ the quintic-loss parameter, $\varepsilon$ the cubic-gain coefficient, and $\beta$ accounting for the effective diffusion (viscosity) or angular filtering [53]. The last term in Eq. (1) represents the effective 2D potential, induced by the transverse modulation of the refractive index in the optical medium.

Following Ref. [23], the initial NRB, with amplitude $A$, mean radius $R$, and width $w$, can be taken (in polar coordinates $r = \sqrt{x^2 + y^2}$ and $\theta$) as

$$u(x, y) = A \mathrm{sech}\left((r-R)/w\right) \cos(N\theta) \exp(ig(x,y)). \quad (2)$$

Here $A$, $R$, and $w$ are the initial amplitude, radius, and radial width of the necklace, respectively. Integer $N$ determines the number of elements ("beads") in the necklace structure, which is $2N$. We stress that, for the NRB described by Eq. (2), each individual "bead" is not a soliton, and adjacent beads differ in phase by $\pi$, hence they repel each other.

The initial phase-modulation profile is described by function $g(x,y)$ in Eq. (2). As a typical example, we take on the following phase profile:

$$g(x,y) = -p\, r[\cos(N\theta)]/k, \quad (3)$$

where $N$ is the same integer as in Eq. (2). This form of $g(x,y)$, which can be readily implemented in the experiment, blends the radial ($\sim r$) and azimuthal, $[\cos(N\theta)]/k$, phase modulations, so as to produce relevant results of the subsequent evolution of the field, which are displayed below. In Eq. (3), $p$ measures the radial gradient of the phase modulation, and $k$ determines its azimuthal anharmonicity. In the absence of the modulation [$p = 0$ in Eq. (3)], the initial necklace set always merges into a single fundamental soliton (see below), i.e., the introduction of the phase modulation is necessary for producing more interesting stable patterns.

## 3. Numerical results

Generic results may be adequately represented for the set of parameters $\delta = 0.5$, $\beta = 0.5$, $\nu = -0.01$, $\mu = -1$, and $\varepsilon = 2.5$ [35], which is considered below. In this case, the amplitude and width of the individual 2D stable fundamental soliton, as found from a numerical solution of Eq. (1) obtained by means of the beam-propagation method, are $A = 1.6$ and $w_{\text{FWHM}} = 2$ [35].

The simulations are performed by means of the split-step Fourier method. Taking into regard that the wave function $u(x, y, z)$ must be well localized for soliton solutions, we have concluded that the domain of size (-25, +25), in both transverse directions, in which the simulations were run, is sufficiently large to avoid effects of the boundaries on the propagation of the solitary waves. The stability of the generated patterns was tested in simulations of Eq. (1), adding a random noise to the initial condition, with the strength amounting to 10% of the soliton's amplitude. In fact, this noise represents strong perturbations, hence persisting structures may be accepted as truly robust ones.

The phase-modulation profile with $k=1$, $p=0.42$, and $N=7$, and the input NRB

with $N=7$ and $R=18$ are shown in Figs. 1(a) and 1(b), respectively. The fourteen "beads" of the NRB coincide with seven "ridges" and seven "troughs" of the phase profile. Note that the "ridges" and "troughs" are characterized, severally, by the tilt toward the center and periphery. Accordingly, the tilted radial phase distribution induces a *radial force*, which pushes the "beads" sitting on the "ridges" and "troughs" outwards and inwards, respectively.

Note that the phase difference between adjacent "beads" is modified by the azimuthal phase modulation, through its depth $p$ and anharmonicity $k$. For this reason, the phase shift between the beads is different from $\pi$, reducing the repulsion between them.

Besides the above-mentioned radial and interaction forces, each "bead" is subject to the action of the viscosity induced by the term $\sim \beta$ in Eq. (1). These three forces reshape the NRB in the course of the evolution.

Typical examples of the evolution of NRBs into stable polygonal soliton clusters are shown in Figs. 2(a)-2(c); for these generic situations, the initial radii of NRBs must be large enough. In these figures and similar ones displayed below, the established patterns show stability and the absence of any residual evolution over extremely long propagation distances z～10000, which amounts to ～100 diffraction lengths corresponding to the overall size of the patetrns.

On the contrary, if the initial NRB radius is small, it evolves into a single stable fundamental soliton, as shown in Figs. 3(a)-3(c). The minimum (threshold) value of the radius, $R_{min}$, necessary for the emergence of the polygonal soliton clusters, is presented in Fig. 4: at $R < R_{\min}$, the NRB inputs fuse into the single fundamental soliton, while if $R \geq R_{\min}$ the NRBs develop the stable polygonal clusters. Figure 4 demonstrates a linear growth of $R_{\min}$ with the increase of $N$.

The increase of radius $R$ of the input allows generating polygonal soliton clusters with larger $N$, i.e., the clusters featuring stronger segmentation into a larger number of solitons. On the other hand, fixing $N$ and increasing the initial radius $R$ to very large values makes borders between individual solitons in the emerging cluster smoother, while keeping its polygonal shape.

The existence of the lower threshold ($R_{\min}$) can be explained as follows. If the initial radius exceeds $R_{\min}$, the interactions of the "beads" in the necklace array become weak, due to the large separation between them (corrections due to next-nearest-neighbor interactions can also be taken into account in this situation, as per Ref. [54]). This argument explains the linear dependence observed in Fig. 4, as $R_{\min} = L_{\min} N / (2\pi)$, where $L_{\min}$ is a characteristic length of the interaction between the fundamental solitons [55]. In this case, the NRBs are affected mainly by the radial and viscosity forces. The former one pushes the "beads" to or away from the center, thus reshaping the necklace into the polygonal array, while the viscosity and weak interaction forces help to maintain the stable configuration of the polygon. Note that each individual element in the necklace patterns observed in Figs. 2(a) and 2(b) features an isotropic (circular) shape, unlike the "beads" in the initial pattern. This is explained by the fact that each element evolves into a stable fundamental soliton.

Next, we study the effect of the anharmonicity parameter $k$ of the initial phase-modulation pattern (3) and radius of the necklace on the evolution of the NRB. First, fixing the radius of the necklace, $R$=16, Fig. 5(a) shows, for $N$=7, domains of different outcomes of the evolution of the phase-modulation profile in the plane of ($p,k$). In this case, we take $k > 1$, while the corresponding results for $k = 1$ (i.e., the harmonic azimuthal profile in Eq. (3)) are more complex, as shown in Fig. 6. When $p$ is too large [e.g., region A in Fig. 5(a)], the initial NRB decays because the "beads" move too rapidly, which gives rise to strong viscous losses, cf. Ref. [56]. If the values of $p$ are large enough but smaller than those corresponding to region A, as in domain B of Fig. 5(a), the NRB evolves into a stable regular heptagonal soliton cluster, through annihilation of a half of the "beads" from the initial set. The partial annihilation also results from the rapid motion of the "beads". When depth $p$ becomes still smaller [region C in Fig. 5(a)], the NRB evolves into a stable quasi-polygonal soliton cluster, preserving the initial number of the "beads". If depth $p$ falls below a critical value [region D in Fig. 5(a)], the "beads" strongly attract each other, which causes the fusion of the NRB into a single fundamental soliton.

Next, we fix the anharmonicity, e.g., $k$=4 in Fig. 5(b) (again, for $N$=7), where regions of the different behavior are shown in the plane of the initial radius $R$ of the necklace and depth $p$. In this situation too, the NRB decays when the depth of the phase-modulation profile is too large, see an example in Fig. 5(c). With the decrease of the depth, the following outcomes of the evolution of the NRB are observed: (i) the formation of a stable heptagonal soliton cluster, through the annihilation of a half of the initial "beads" [Fig. 5(d)]; (ii) the transition to a stable quasi-heptagonal soliton cluster, which preserves the initial number of the "beads" [Fig. 5(e)], and (iii) the merger into a single fundamental soliton [Fig. 5(f)].

As mentioned above, the dynamics of the NRB modulated by the imprinted phase profile is more complex in the case of $k=1$. Figure 6(a) shows the diagram of different dynamical regimes in the plane of initial radius $R$ and depth $p$ for $N=7$. Similar to the results shown in Fig. 5(b), when the depth $p$ changes from large to small values, the dynamics of the NRB varies through the sequence including (i) the decay of the cluster [Fig. 6(b)], (ii) the formation of a stable heptagonal soliton cluster with the annihilation of the half of the initial "beads" [Fig. 6(c)], (iii) the generation of a stable *quasi-heptagonal soliton cluster*, which preserves the initial number of the "beads" [Fig. 6(d)], and (iv) the merger into a fundamental soliton [Fig. 6(e)]. However, when the depth of the phase-modulation profiles becomes still smaller, the NRB evolves into a stable *regular heptagonal soliton cluster,* which preserves the initial number of the "beads" [Fig. 6(f)] (recall that patterns of this type were produced with $k > 1$). Lastly, if depth $p$ is too small, the NRB merges into a single fundamental soliton [Fig. 6(g)].

In addition, to check if the viscous term in the CQ CGL equation is necessary for the generation of these varieties of the soliton clusters, we have also simulated the dynamics of the NRB with the imprinted phase-modulation profile without this term. For instance, with typical values of the parameters chosen in addition to $\beta = 0$, *viz.*, $\delta = 0.5$, $\nu = -0.2$, $\mu = -1$, and $\varepsilon = 1.6$, the NRB quickly collapses in the absence of the viscosity (actually, regardless of particular values of the parameters of the phase modulation), see Fig. 7. Thus, the viscous term [$\beta \neq 0$ in Eq. (1)] is necessary to support the dynamical scenarios outlined above.

As mentioned above, the CQ nonlinearity in the CGL equation models the saturation of the inner gain in the optical medium. It is known that the nonlinearity featuring the explicit saturation leads to essentially the same results as the CQ model (see, e.g., Refs. [37-40] and [47]).

Figure 8 demonstrates that the replacement of the CQ terms in Eq. (1) by the full saturable nonlinearity, $(1-i\eta)|u|^2 u/(1+\gamma|u|^2)$, does not essentially change the results in the present context either. In this case, fixing, e.g., $R=16$, $N=5$, and $k=1$, we observe the decay of the NRB at $p \geq 0.8$ [Fig. 8(a)], the generation of a stable pentagonal soliton cluster, with the annihilation of a half of the initial "beads", at $0.5 < p < 0.8$ [Fig. 8(b)], the generation of a stable regular pentagonal soliton cluster at $0.1 < p \leq 0.5$ [Fig. 8(c)], and the fusion into a stable fundamental soliton at $p \leq 0.1$ [Fig. 8(d)]. Thus, the basic dynamical scenarios outlined above are quite universal ones, which are not drastically affected by the particular form of the nonlinearity in the CGL.

## 4. Conclusions

In this work, we have studied the dynamics of the necklace-ring beams (NRBs) carrying the initial radial-azimuthal phase modulation, in the framework of the 2D CQ (cubic-quintic) CGL (complex Ginzburg-Landau) model with the viscosity term. We have found that the necklace carrying such phase modulation profiles can evolve into three distinct stable patterns: (i) regular polygonal soliton clusters, (b) quasi-polygonal soliton clusters, and (c) single fundamental solitons. These generic outcomes can be controlled via varying the depth and anharmonicity of the initial phase-modulation profile [Eq. (3)], or the radius and the number of "beads" in the initial necklace set [Eq. (2)]. Extensive numerical simulations have been performed to identify parameter domains for the formation of a variety of stable patterns from the initial necklace patterns. Threshold values of the initial phase-modulation depth and radius of the necklace ring, which determine boundaries between different outcomes of the evolution, have been found. It has been demonstrated too that the model with the saturable nonlinearity yields essentially the same results as its counterpart with the CQ terms; hence the basic dynamical scenarios reported above are universal for the CGL equations. We did not consider NRB inputs with embedded vorticity, as the presence of the viscosity does not allow rotation of patterns.

A challenging possibility is to extend the analysis to the three-dimensional case, i.e., to the study of the generation of complex *spatiotemporal* soliton patterns [57].

**Acknowledgment**

This work was supported by the National Natural Science Foundation of China (Grant No. 11174061) and the Guangdong Province Natural Science Foundation of China (Grant No. S2011010005471). The work of D.M. was supported by CNCS-UEFISCDI, project number PN-II-ID-PCE-2011-3-0083.

**References**

[1] *Spatial Solitons*, ed. by S. Trillo and W. Torruellas (Springer: Berlin, 2001).

[2] Y. S. Kivshar and G. P. Agrawal, *Optical Solitons: From Fibers to Photonic Crystals* (Academic Press: San Diego, 2003).

[3] W. J. Firth, in *Self-Organization in Optical Systems and Applications in Information Technology*, edited by M. A. Vorontsov and W. B. Miller (Springer-Verlag, Berlin, 1995), p. 69.

[4] N. N. Rosanov, S.V. Fedorov, A.N. Shatsev, In: Dissipative Solitons: from Optics to Biology and Medicine, N. Akhmediev and A. Ankiewicz, Eds., Lecture Notes in Physics, **751** 93-111 (2008).

[5] R. Kuszelewicz, S. Barbay, G. Tissoni, and G. Almuneau, Eur. Phys. J. D **59**, 1 (2010).

[6] E. Pampaloni, P. L. Ramazza, S. Residori, and F. T. Arecchi, Phys. Rev. Lett. **74**, 258 (1995).

[7] W. J. Firth and A. J. Scroggie, Phys. Rev. Lett. **76**, 1623 (1996).

[8] L. Spinelli, G. Tissoni, M. Brambilla, F. Prati, and L. A. Lugiato, Phys. Rev. A **58**, 2542 (1998).

[9] Z. Chen and K. McCarthy, Opt. Lett. **27**, 2019 (2002).

[10] S. Barland, J. R. Tredicce, M. Brambilla, L. A. Lugiato, S. Balle, M. Giudici, T. Maggipinto, L. Spinelli, G. Tissoni, T. Knödl, M. Miller, and R. Jäger, Nature (London) **419**, 699 (2002).

[11] J. W. Fleischer, M. Segev, N. K. Efremidis, and D. N. Christodoulides, Nature

(London) **422**, 147 (2003).

[12] Y. V. Kartashov, A. A. Egorov, L. Torner, and D. N. Christodoulides, Opt. Lett. **29**, 1918 (2004).

[13] B. Freedman, G. Bartal, M. Segev, R. Lifshitz, D. N. Christodoulides, and J. W. Fleischer, Nature (London) **440**, 1166 (2006).

[14] P. Genevet, S. Barland, M. Giudici, and J. R. Tredicce, Phys. Rev. Lett. **104**, 223902 (2010).

[15] C. Cleff, B. Gütlich, and C. Denz, Phys. Rev. Lett. **100**, 233902 (2008).

[16] Z. Chen, H. Martin, E. D. Eugenieva, J. Xu, and J. Yang, Opt. Express **13**, 1816 (2005).

[17] A. S. Desyatnikov and Yu. S. Kivshar, Phys. Rev. Lett. **88**, 053901 (2002).

[18] Y. V. Kartashov, L.-C. Crasovan, D. Mihalache, and L. Torner, Phys. Rev. Lett. **89**, 273902 (2002).

[19] J. J. Garcia-Ripoll, V. M. Perez-Garcia, E. A. Ostrovskaya, and Y. S. Kivshar, Phys. Rev. Lett. **85**, 82 (2000).

[20] D. Neshev, W. Krolikowski, D. E. Pelinovsky, G. McCarthy, and Y. S. Kivshar, Phys. Rev. Lett. **87**, 103903 (2001).

[21] T. Carmon, R. Uzdin, C. Pigier, Z. H. Musslimani, M. Segev, and A. Nepomnyashchy, Phys. Rev. Lett. **87**, 143901 (2001).

[22] Y. V. Kartashov, R. Carretero-González, B. A. Malomed, V. A. Vysloukh, and L. Torner, Opt. Express **13**, 10703 (2005).

[23] M. Soljăcić, S. Sears, and M. Segev, Phys. Rev. Lett. **81**, 4851 (1998); *ibid*. **86**,

420 (2001).

[24] D. Mihalache, D. Mazilu, L.-C. Crasovan, B. A. Malomed, F. Lederer, and L. Torner, Phys. Rev. E **68**, 046612 (2003).

[25] J. Yang, I. Makasyuk, P. G. Kevrekidis, H. Martin, B. A. Malomed, D. J. Frantzeskakis, and Z. Chen, Phys. Rev. Lett. **94**, 113902 (2005).

[26] N. N. Akhmediev, V. V. Afanasjev, and J. M. Soto-Crespo, Phys. Rev. E **53**, 1190 (1996).

[27] W. H. Renninger, A. Chong, and F. W. Wise, Phys. Rev. A **77**, 023814 (2008).

[28] N. Akhmediev and A. Ankiewicz, *Dissipative Solitons*, Vol. 661 of Lecture Notes in Physics (Springer, 2005).

[29] H. Leblond, B. A. Malomed, and D. Mihalache, Phys. Rev. A **80**, 033835 (2009).

[30] L.-C. Crasovan, B. A. Malomed, and D. Mihalache, Phys. Rev. E **63,** 016605 (2001).

[31] L.-C. Crasovan, B. A. Malomed, and D. Mihalache, Phys. Lett. A **289**, 59 (2001).

[32] D. V. Skryabin and A. G. Vladimirov, Phys. Rev. Lett. **89**, 044101 (2002).

[33] D. Mihalache, D. Mazilu, F. Lederer, Y. V. Kartashov, L.-C. Crasovan, L. Torner, and B. A. Malomed, Phys. Rev. Lett. **97,** 073904 (2006).

[34] D. Mihalache, D. Mazilu, F. Lederer, H. Leblond, and B. A. Malomed, Phys. Rev. A **75,** 033811 (2007).

[35] Y. J. He, B. A. Malomed, and H. Z. Wang, Opt. Express **15**, 17502 (2007).

[36] V. I. Petviashvili and A. M. Sergeev, Dokl. AN SSSR **276**, 1380 (1984) [English translation: Sov. Phys. Doklady **29**, 493 (1984)].

[37] B. A. Malomed, Complex Ginzburg-Landau equation in: Encyclopedia of Nonlinear Science, A. Scott, ed., (Routledge, New York, 2005) pp. 157-160.

[38] B. A. Malomed, Chaos **17**, 037117 (2007).

[39] P. Mandel and M. Tlidi, J. Opt. B: Quantum Semiclassical Opt. **6**, R60 (2004).

[40] N. N. Rosanov, S. V. Fedorov, and A. N. Shatsev, Appl. Phys. B: Lasers Opt. **81**, 937 (2005).

[41] V. Skarka and N. B. Aleksić, Phys. Rev. Lett. **96**, 013903 (2006).

[42] N. Akhmediev, J. M. Soto-Crespo, and Ph. Grelu, Chaos **17**, 037112 (2007).

[43] V. Skarka, N. B. Aleksić, H. Leblond, B. A. Malomed, and D. Mihalache, Phys. Rev. Lett. **105**, 213901 (2010).

[44] D. Mihalache, Proc. Romanian Acad. A **11**, 142 (2010); D. Mihalache, Rom. J. Phys. **57**, 352 (2012).

[45] D. Mihalache, Rom. Rep. Phys. **63**, 9 (2011); D. Mihalache, D. Mazilu, Rom. Rep. Phys. **60**, 749 (2008); D. Mihalache, Rom. Rep. Phys. **63,** 325 (2011).

[46] Y. V. Kartashov, V. V. Konotop, and V. A. Vysloukh, Opt. Lett. **36**, 82 (2011).

[47] P. V. Paulau, D. Gomila, P. Colet, B. A. Malomed, and W. J. Firth, Phys. Rev. E **84**, 036213 (2011).

[48] Y. J. He, B. A. Malomed, F. Ye, J. Dong, Z. Qiu, H. Z. Wang, and B. Hu, Phys. Scr. **82,** 065404 (2010).

[49] Y. J. He, B. A. Malomed, D. Mihalache, F. Ye, and B. Hu, J. Opt. Soc. Am. B **27**, 1266 (2010).

[50] M. Tlidi, P. Mandel, and R. Lefever, Phys.Rev.Lett. **73,** 640 (1994).

[51] M. Tlidi, A. G. Vladimirov, and P. Mandel, IEEE J. Quant. Electron. **39**, 216 (2003).

[52] A. G. Vladimirov, R. Lefever, and M. Tlidi, Phys. Rev. A **84**, 043848 (2011).

[53] J. M. Soto-Crespo, N. Akhmediev, C. Mejía-Cortés, and N. Devine, Opt. Express **17**, 4236 (2009).

[54] A. G. Vladimirov, J. M. McSloy, D. V. Skryabin, and W. J. Firth, Phys. Rev. E **65**, 046606 (2002).

[55] B. A. Malomed, Phys. Rev. E **58**, 7928 (1998).

[56] Y. J. He, B. A. Malomed, D. Mihalache, B. Liu, H. C. Huang, H. Yang, and H. Z. Wang, Opt. Lett. **34**, 2976 (2009).

[57] B. A. Malomed, D. Mihalache, F. Wise, and L. Torner, J. Opt. B: Quantum Semiclass. Opt. **7**, R53 (2005).

# Figure captions

Fig. 1. (Color online) (a) The initial radial-azimuthal phase-modulation profile for $k=1$, $p=0.42$, and $N=7$. (b) The initial necklace ring with $N=7$ and $R=18$.

Fig. 2. (Color online) The evolution of necklace rings into various stable polygonal soliton clusters for $k=1$. (a) A pentagonal cluster for $N=5$, $R=16$, and $p=0.6$. (b) A hexagonal cluster for $N=6$, $R=17$, and $p=0.45$. (c) A heptagonal cluster for $N=7$, $R=18$, $p=0.42$. Here and in all other figures, the transverse domain is $(-25,+25) \times (-25,+25)$.

Fig. 3. (Color online) The fusion of NRBs into stable fundamental solitons for $k=1$: (a) $N=5$, $R=8$, and $p=0.6$; (b) $N=6$, $R=9$, $p=0.45$; (c) $N=7$, $R=10$, and $p=0.42$.

Fig. 4. The dependence of the threshold value of the initial radius $R_{min}$ of the NRB on its integer azimuthal index $N$. The initial radius must exceed $R_{min}$ to let the input form stable polygonal soliton clusters.

Fig. 5. (Color online) Regions of depth $p$ of the phase-modulation profile (3) for $N=7$ versus (a) anharmonicity parameter $k$ for $R=16$, and (b) the radius of the input NRB for $k=4$. In (a) and (b), the areas are labeled as follows. A: The decay of the NRB; B: the self-trapping into a stable heptagonal soliton cluster through annihilation of a half of the initial "beads"; C: the evolution into a stable quasi-heptagonal soliton cluster preserving the initial number of "beads"; D: the fusion of the NRB into a stable fundamental soliton. (c) An example of the decay of the initial NRB for $p=6$ and $k=4$. (d) The generation of a stable

heptagonal soliton cluster through the annihilation of a half of the "beads" for $p=4.6$ and $k=4$. (e) The generation of a stable quasi-heptagonal soliton cluster for $p=2.5$ and $k=4$. (f) The generation of the single fundamental soliton for $p=0.5$ and $k=4$.

Fig. 6. (Color online) (a) Regions of the different behavior in the plane of ($R$,$p$) for $N=7$ and $k=1$. A: The decay of the NRB; B: the evolution of the NRB into a stable heptagonal soliton cluster with the annihilation of a half of the initial "beads"; C: the evolution of the NRB into a stable quasi-heptagonal soliton cluster, which preserves the initial number of the "beads"; E: the evolution of the NRB into a stable regular heptagonal soliton cluster; D and F: the merger into the single fundamental soliton. (b) An example of the decay of the NRB for $R=17$ and $p=1.4$. (c) The generation of a stable heptagonal soliton cluster, with the annihilation of a half of the initial "beads" for $R=17$ and $p=1.2$. (d) The generation of a stable quasi-heptagonal soliton cluster preserving the initial number of the "beads" for $R=17$ and $p=1.0$. (e) The generation of a stable fundamental soliton for $R=17$ and $p=0.48$. (f) The generation of a stable regular heptagonal soliton cluster for $R=17$ and $p=0.44$. (g) The generation of a stable fundamental soliton for $R=17$ and $p=0.3$.

Fig. 7. (Color online) The collapse of the NRB in the case of $\beta = 0$, for $R=17$, $N=7$, $k=1$, and $p=0.44$ (in the absence of the vicosity).

Fig. 8. (Color online) Typical evolution scenarios of the NRB in the model with the cubic-quintic terms in Eq. (1) replaced by the saturable nonlinearity,

$(1-i\eta)|u|^2 u / (1+\gamma|u|^2)$ (here, $\eta = 2.85$, and $\gamma = 0.5$). (a) An example of the decay of the NRB at *p*=0.8. (b) The generation of a stable pentagonal soliton cluster with the annihilation of a half of the initial "beads" at *p*=0.6. (c) The generation of a stable regular pentagonal soliton cluster at *p*=0.5. (g) The generation of a stable fundamental soliton at *p*=0.1. Other parameters are *R*=16, *N*=5, and *k*=1.

**Figures**

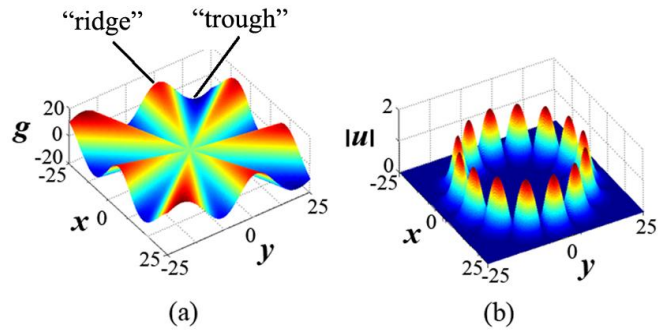

(a)  (b)

Fig. 1. (Color online) (a) The initial radial-azimuthal phase-modulation profile for $k=1$, $p=0.42$, and $N=7$. (b) The initial necklace ring with $N=7$ and $R=18$.

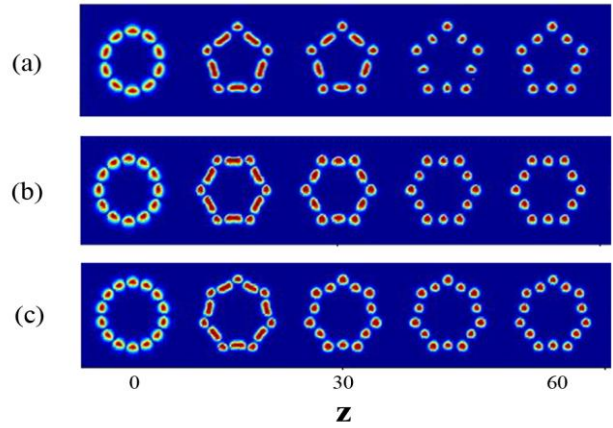

Fig. 2. (Color online) The evolution of necklace rings into various stable polygonal soliton clusters for $k=1$. (a) A pentagonal cluster for $N=5$, $R=16$, and $p=0.6$. (b) A hexagonal cluster for $N=6$, $R=17$, and $p=0.45$. (c) A heptagonal cluster for $N=7$, $R=18$, $p=0.42$. Here and in all other figures, the transverse domain is $(-25,+25) \times (-25,+25)$.

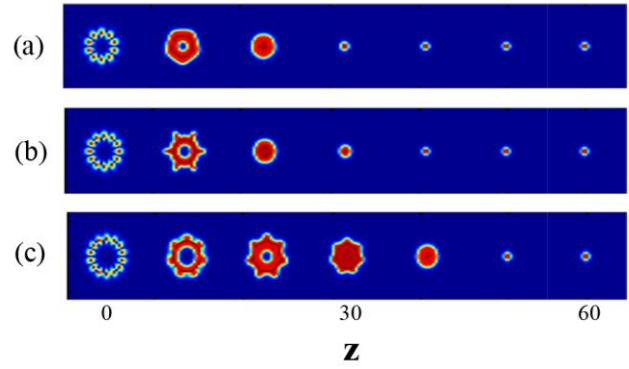

Fig. 3. (Color online) The fusion of NRBs into stable fundamental solitons for $k=1$: (a) $N=5$, $R=8$, and $p=0.6$; (b) $N=6$, $R=9$, $p=0.45$; (c) $N=7$, $R=10$, and $p=0.42$.

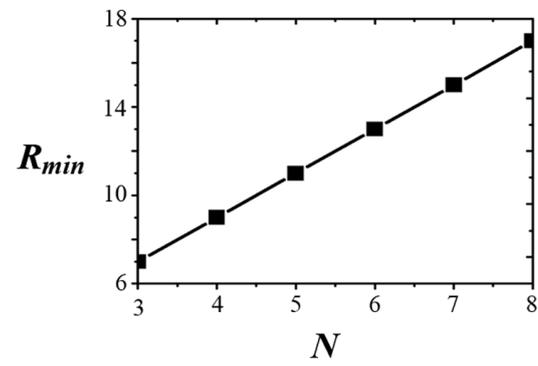

Fig. 4. The dependence of the threshold value of the initial radius $R_{min}$ of the NRB on its integer azimuthal index $N$. The initial radius must exceed $R_{min}$ to let the input form stable polygonal soliton clusters.

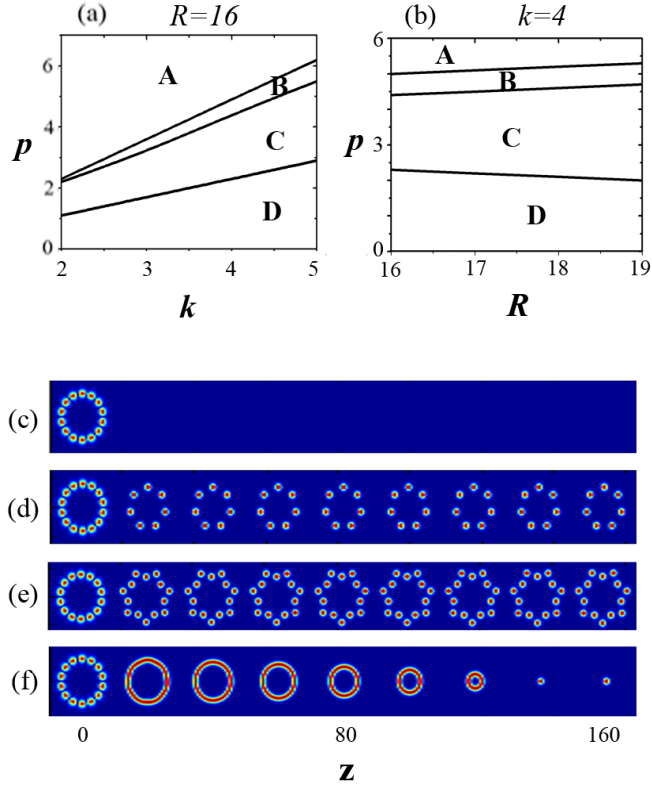

Fig. 5. (Color online) Regions of depth $p$ of RAPM profile for $N=7$ versus (a) the anharmonicity parameter $k$ of the RAPM profile for $R=16$, and (b) the radius of the input NRB for $k=4$. In (a) and (b) the areas are labeled as follows. A: Decay of the NRB; B: self-trapping into a stable heptagonal soliton cluster through annihilation of a half of the initial "beads"; C: the evolution into a stable quasi-heptagonal soliton cluster preserving the initial number of "beads"; D: the merger of the NRB into a stable fundamental soliton. (c) An example of the decay of the initial NRB for $p=6$ and $k=4$; (d) The generation of a stable heptagonal soliton cluster through the annihilation of a half of the "beads" for $p=4.6$ and $k=4$; (e) The generation of a stable quasi-heptagonal soliton cluster for $p=2.5$ and $k=4$; (f) The generation of the single fundamental soliton for $p=0.5$ and $k=4$.

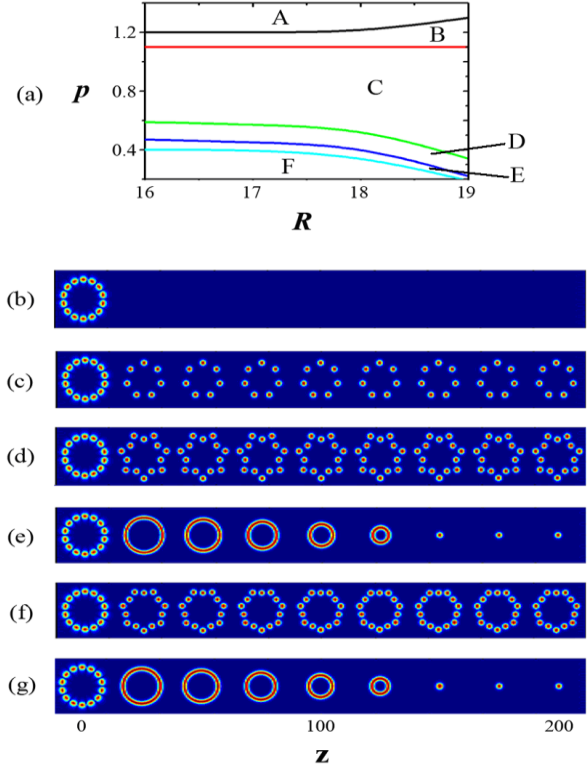

Fig. 6. (Color online) (a) Regions of the different behavior in the plane of ($R$, $p$) for $N=7$ and $k=1$. A: The decay of the NRB; B: the evolution of the NRB into a stable heptagonal soliton cluster with the annihilation of a half of the initial "beads"; C: the evolution of the NRB into a stable quasi-heptagonal soliton cluster, which preserves the initial number of the "beads"; E: the evolution of the NRB into a stable regular heptagonal soliton cluster; D and F: the merger into the single fundamental soliton. (b) An example of the decay of the NRB for $R=17$ and $p=1.4$. (c) The generation of a stable heptagonal soliton cluster, with the annihilation of a half of the initial "beads" for $R=17$ and $p=1.2$. (d) The generation of a stable quasi-heptagonal soliton cluster preserving the initial number of the "beads" for $R=17$ and $p=1.0$. (e) The generation of a stable fundamental soliton for $R=17$ and $p=0.48$; (f) The generation of a stable regular heptagonal soliton cluster for $R=17$ and $p=0.44$. (g) The generation of a stable fundamental soliton for $R=17$ and $p=0.3$.

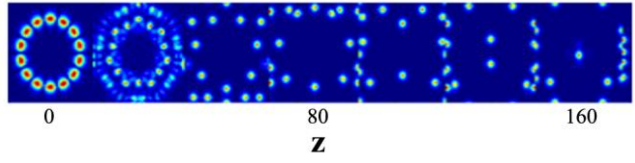

Fig. 7. (Color online) The collapse of the NRB in the case of $\beta = 0$, for *R*=17, *N*=7, *k*=1, and *p*=0.44 (in the absence of the vicosity).

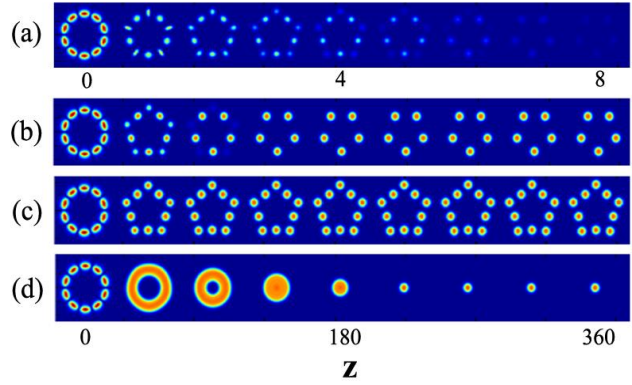

Fig. 8. (Color online) Typical evolution scenarios of the NRB in the model with the cubic-quintic terms in Eq. (1) replaced by the saturable nonlinearity, $(1-i\eta)|u|^2 u/(1+\gamma|u|^2)$ (here, $\eta=2.85$, and $\gamma=0.5$). (a) An example of the decay of the NRB at $p=0.8$. (b) The generation of a stable pentagonal soliton cluster with the annihilation of a half of the initial "beads" at $p=0.6$. (c) The generation of a stable regular pentagonal soliton cluster at $p=0.5$. (g) The generation of a stable fundamental soliton at $p=0.1$. Other parameters are $R=16$, $N=5$, and $k=1$.